\documentclass[pra,floatfix,11pt,superscriptaddress,reprint]{revtex4-1}
\usepackage[utf8]{inputenc}
\usepackage{hyperref}
\usepackage{graphicx}

\def\br{\mathbf{r}}
\def\bk{\mathbf{k}}
\def\lENZ{\lambda_{\mathrm{ENZ}}}

\begin{document}

\title{Two-beam coupling by a hot electron nonlinearity}

\author{J. Paul}
\affiliation{Physical Measurement Laboratory, National Institute of Standards and Technology, Gaithersburg, MD 20899, USA}
\author{M. Miscuglio}
\affiliation{Department of Electrical and Computer Engineering, George Washington University, Washington, DC 20052, USA}
\author{Y. Gui}
\affiliation{Department of Electrical and Computer Engineering, George Washington University, Washington, DC 20052, USA}
\author{V. J. Sorger}
\affiliation{Department of Electrical and Computer Engineering, George Washington University, Washington, DC 20052, USA}
\author{J. K. Wahlstrand}
\email{jared.wahlstrand@nist.gov}
\affiliation{Physical Measurement Laboratory, National Institute of Standards and Technology, Gaithersburg, MD 20899, USA}

\begin{abstract}
Transparent conductive oxides such as indium tin oxide (ITO) bear the potential to deliver efficient all-optical functionality due to their record-breaking optical nonlinearity at epsilon near zero (ENZ) wavelengths. All-optical applications generally involve more than one beam, but the coherent interaction between beams has not previously been discussed in materials with a hot electron nonlinearity. Here we study the optical nonlinearity at ENZ in ITO and show that spatial and temporal interference has important consequences in a two beam geometry. Our pump-probe results reveal a polarization-dependent transient that is explained by momentary diffraction of pump light into the probe direction by a temperature grating produced by pump-probe interference. We further show that this effect allows tailoring the nonlinearity by tuning frequency or chirp. Having fine control over the strong and ultrafast ENZ nonlinearity may enable applications in all-optical neural networks, nanophotonics, and spectroscopy. 
\end{abstract}

\maketitle

Recent years have seen growing interest in the nonlinear optics of transparent conductive oxides (TCO) such as indium tin oxide (ITO) and aluminum zinc oxide \cite{alam_large_2016,caspani_enhanced_2016,clerici_controlling_2017,carnemolla_degenerate_2018,alam_large_2018,kinsey_nonlinear_2019,reshef_nonlinear_2019,bruno_negative_2020}.
Resonant enhancement of the nonlinearity at wavelengths where the dielectric function epsilon is near zero (ENZ) enables large effects at subwavelength interaction lengths \cite{alam_large_2016,caspani_enhanced_2016,carnemolla_degenerate_2018}.
The addition of a nanoparticle array results in excitation of a localized surface plasmon resonance and an even higher nonlinearity \cite{alam_large_2018,bruno_negative_2020}.
Kinsey and Khurgin have emphasized that ENZ materials have two parameters that can be tuned: the effective nonlinear susceptibility $\chi^{(3)}$, which provides the nonlinear response, and the linear dielectric function $\varepsilon$, which determines the magnitude and wavelength range of the ENZ enhancement \cite{kinsey_nonlinear_2019}.
In general, many physical processes can contribute to the effective $\chi^{(3)}$.
The dominant effect in ITO for near infrared pulses is thought to be a ``hot electron nonlinearity,'' which results from optical heating of electrons combined with an electron temperature-dependent linear optical response \cite{rotenberg_nonlinear_2007,conforti_derivation_2012,boyd_third-order_2014,alam_large_2016}.
The electron temperature builds up during the laser pulse, then falls rapidly as the electrons interact with the lattice.
In ITO, the cooling takes place over hundreds of femtoseconds.
To lowest order, the temperature change is linear in intensity, causing an effective third-order nonlinearity consisting of a refractive component, characterized by the Kerr coefficient $n_2$, and an absorptive component, characterized by the two-photon absorption coefficient $\beta$.
In ITO near ENZ, these coefficients have record-setting magnitude \cite{reshef_nonlinear_2019}.

\begin{figure}
    \centering
    \includegraphics[width=8cm]{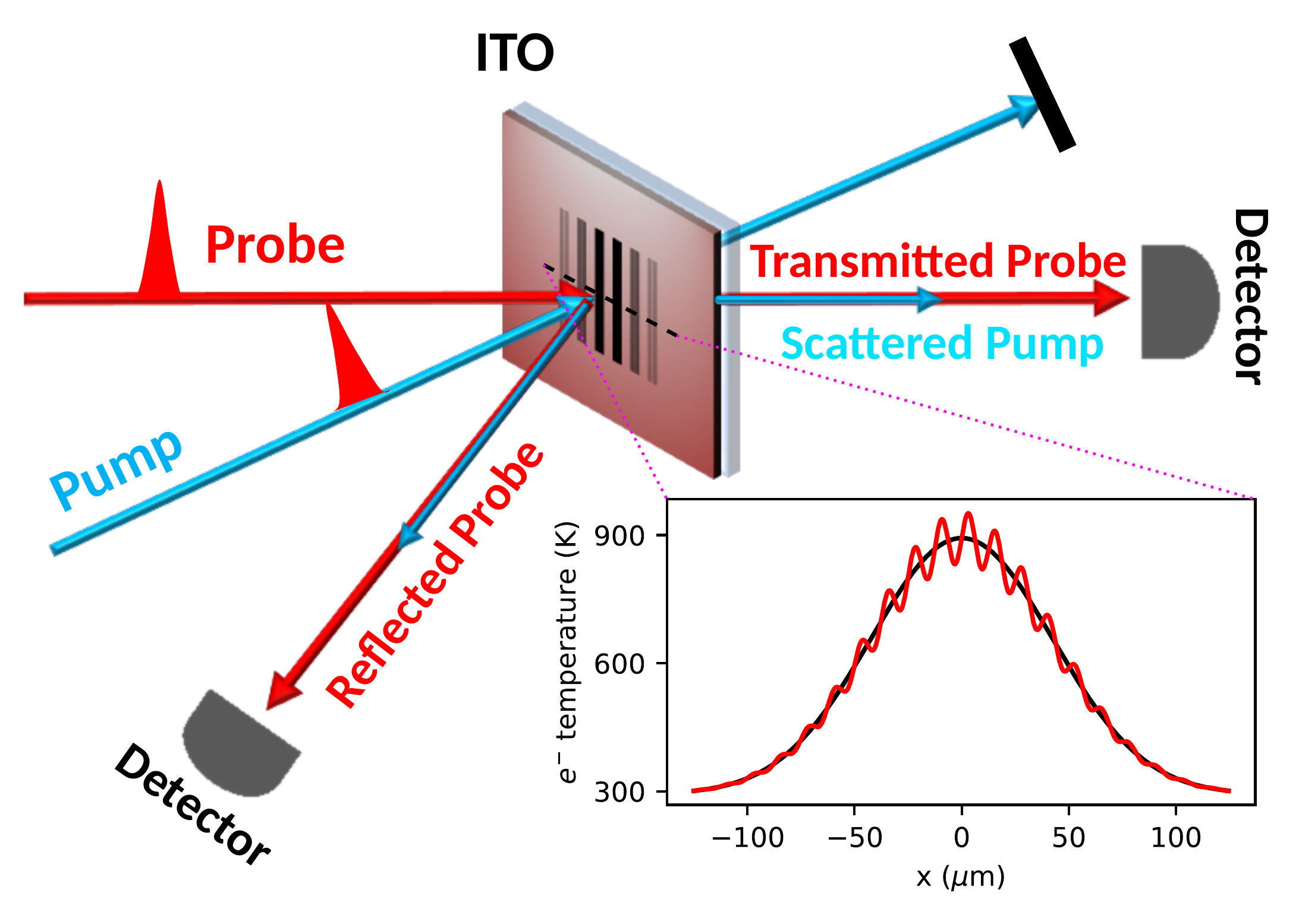}
    \caption{Diagram illustrating TBC in a material with a hot electron nonlinearity. Pump and probe beams that overlap in time, space, and polarization state interfere, resulting in a periodic modulation of the electron temperature in the film. The inset shows temperature along a slice inside the film for co-polarized (red) and cross polarized pulses (black). The temperature grating results in diffraction of the pump beam into the reflected and transmitted probe direction. The effect only appears when the two pulses overlap in time and when they share a polarization component.}
    \label{intro_fig}
\end{figure}

While the hot electron nonlinearity is thought to be the most important contributor to $\chi^{(3)}$ in ITO at ENZ wavelengths, nonlinear optics in conductive films is complicated.
Other effects may contribute, such as interband excitation of carriers or a bound electronic nonlinearity.
The strength of the various contributors depends on wavelength, pulse duration, and material properties \cite{boyd_third-order_2014,clerici_controlling_2017}.
It is possible to separate various contributors to $\chi^{(3)}$ using pump-probe measurements \cite{wahlstrand_effect_2013,reichert_temporal_2014,wahlstrand_absolute_2015}.
However, doing so quantitatively requires accounting for coherent coupling between pulses when they overlap in time.
This coupling arises from a nonlinear transient grating created by interference of the two beams.
As shown pictorially in Fig.~\ref{intro_fig}, the nonlinear grating diffracts pump light into the probe direction, changing the effective nonlinearity for the probe beam.
The effect of the grating depends on the temporal and spectral dependence of the underlying nonlinearity and generally requires modeling.
Transient nonlinear diffraction or \emph{two-beam coupling} (TBC) has been studied in many contexts \cite{silberberg_instabilities_1982,dogariu_purely_1997,tang_time-domain_1997,smolorz_femtosecond_2000,bernstein_two-beam_2009,wahlstrand_effect_2011,wahlstrand_effect_2013,michel_dynamic_2014}.
However, to our knowledge it has not been discussed in the context of ENZ nonlinear optics, where the refractive and absorptive components of the susceptibility are often of similar magnitude.

Here we present results of pump-probe experiments on an ITO film and numerical simulations of TBC.
First, we show that TBC produces a polarization dependent change in transmission that follows the pump intensity envelope, mimicking an instantaneous nonlinearity resulting from a bound electronic $\chi^{(3)}$.
When we calculate TBC using a simple model, we find that the magnitude of the observed polarization anisotropy can be explained by the hot electron nonlinearity, with no need for another source of effective $\chi^{(3)}$.
This provides a physical interpretation for the degenerate nonlinearity in TCO's, the subject of a recent paper \cite{carnemolla_degenerate_2018}.
Second, we show that the time dependence of the hot electron nonlinearity in TCO's has an important consequence for a slightly nondegenerate two-beam experiment.
The grating can effectively mix the refractive response into the absorptive response and vice versa, potentially an important tool for tailoring the nonlinear interaction, with implications in applications of TCO's for all-optical devices.

\begin{figure}[t]
    \centering
    \includegraphics[width=7.5cm]{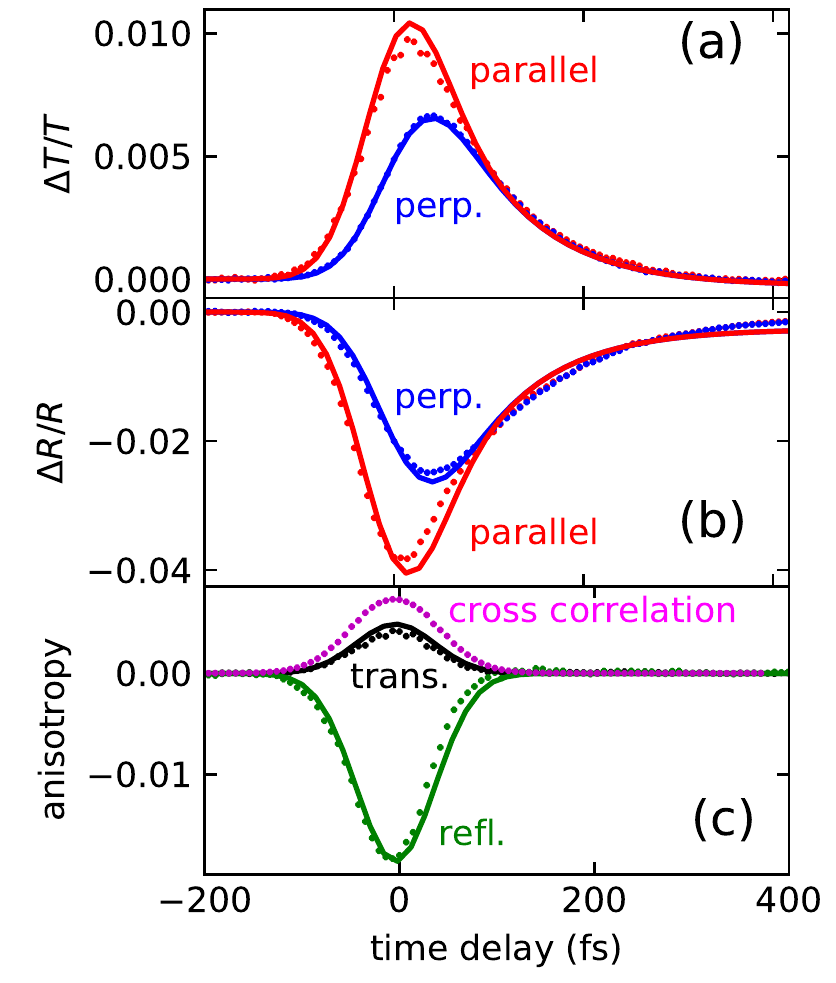}
    \caption{Experimental pump-probe traces (points) and theory curves (lines) for transform-limited pulses at near normal incidence in ITO. Data is shown for parallel (red) and perpendicular (blue) polarization. (a) Change in probe transmission. (b) Change in probe reflection. (c) Transient birefringence for transmission (black) and reflection (green). The normalized cross correlation between the pump and probe beams is shown for comparison as magenta dots.}
    \label{fig:pp1}
\end{figure}

We perform pump-probe experiments on a 316 nm thick ITO film.
The light source is an optical parametric amplifier (OPA) operating at 1 MHz repetition rate.
The ENZ wavelength $\lENZ$ of the sample is 1240 nm, and the laser wavelength is centered there for all experiments.
All experiments were performed at 0$^\circ$ incidence angle.
The maximum pump intensity used is approximately 4 GW/cm$^2$.
More details of the experiment are provided in the supplemental document.
Measured pump-induced changes in probe transmission and reflection are shown in Fig.~\ref{fig:pp1}.
The change in transmission (Fig.~\ref{fig:pp1}a) and reflection (Fig.~\ref{fig:pp1}b) are shown for probe polarization parallel (red dots) and perpendicular (blue dots) to the pump polarization.
For perpendicular polarization, we measure a transmission increase that accumulates during the pump pulse, followed by an exponential decay, with a time constant of approximately 85 fs.
We observe a similar decrease in the probe reflectivity.
This is the laser-induced electron heating and rapid cooling response previously measured in ITO  \cite{alam_large_2016,wang_extended_2019}.
For parallel polarization, we observe an additional transient increase (decrease) in transmission (reflection) for time delays during the pump pulse, but at longer delay the signal drops to the same level as for perpendicular polarization.
The polarization anisotropy, found by subtracting the perpendicular probe data from the parallel probe data, isolates this fast transient effect and is shown in Fig.~\ref{fig:pp1}c.
For comparison, the measured cross correlation is also shown in Fig.~\ref{fig:pp1}c.
The anisotropy in both transmission and reflectivity follows the pump intensity envelope convolved with the probe intensity envelope.

A $\chi^{(3)}$ response that is truly third-order in the optical field produces a polarization-dependent effect that rises and falls with the pump intensity envelope, consistent with the observed transient birefringence.
However, the band gap of ITO is approximately 3.5 eV, so for the 1 eV photon energy used here, two-photon absorption is not allowed and one would expect $\mathrm{Im}[\chi^{(3)}] = 0$ and thus $\beta = 0$.
A purely real $\chi^{(3)}$ would not produce a significant change in transmission.
Instead we shall show that TBC explains the observed polarization dependence.

Following previous work, we calculate TBC for pulses in the slowly-varying envelope approximation \cite{smolorz_femtosecond_2000,wahlstrand_effect_2013}.
For probe intensity much weaker than the pump intensity, the total intensity is approximately
\begin{equation}
I(\br,t) \approx \frac{n_0c}{8\pi} \left[ |A_e(t)|^2 + \left(A_e^*(t) A_x(t-t_d) e^{i\Delta \bk \cdot \br-i\delta t} + c.c. \right)\right],
\label{intensity_grating}
\end{equation}
where $n_0$ is the linear refractive index, $A_e(t)$ is the complex pump envelope, $A_x(t-t_d)$ is the complex probe envelope along the pump polarization direction (delayed with respect to the pump by $t_d$), $\Delta \bk$ is the wavevector difference between pump and probe beams, and $\delta$ is the carrier frequency difference.
The first term in brackets, which we call the ``smooth'' term, is the intensity of the pump beam alone.
The second term, the ``grating'' term, is due to interference of the pump and probe beams and is nonzero only when $t_d$ is near zero.

To calculate TBC we need a model of the hot electron nonlinearity.
We first use a phenomenological two-temperature model to calculate the time-dependent electron temperature $T_e$ and lattice temperature $T_l$ \cite{rotenberg_nonlinear_2007,conforti_derivation_2012}.
We assume uniform heating throughout the depth of the film.
The equations for temperature are $C_e dT_e/dt =-g_{el} (T_e - T_l) + S(t)$, and $C_l dT_l/dt = g_{el} (T_e - T_l)$, where $g_{el}$ is an electron-lattice coupling parameter, $C_e$ and $C_l$ are the heat capacities of the electron gas and lattice, respectively, and $S(t) = \eta I(t)$ is a heating term.
We calculate smooth and grating terms analogous to those in Eq.~(\ref{intensity_grating}) such that we can express the time-dependent electron temperature inside the pump spot as $T_e(\br,t) = T^s_e(t) + [T^g_e(t) e^{i\bk \cdot \br}+c.c.]$.
To model the dependence of the complex dielectric function on $T_e$, we use the modified Drude model developed by Wang \emph{et al.} for ITO \cite{wang_extended_2019}.
For our calculations, we use the parameters reported in \cite{wang_extended_2019}, except we force the change in dielectric function to be linear at low intensity.
For more details, see the supplemental document.

Using $T_e^s$ and $T_e^g$, we can straightforwardly separate the dielectric function into smooth and grating terms $\varepsilon^s$ and $\varepsilon^g$.
Using a standard transfer matrix calculation, the outgoing field in the probe direction is calculated, from which changes in transmission and reflection are calculated.
The grating term corresponds to diffracted pump light interfering with the outgoing probe field, resulting in an additional amplitude change and phase shift.
The effect does not rely on the probe beam being intense; the amplitude of the grating scales with the probe intensity, and therefore so does the amount of pump light scattered into the probe direction.
The changes in probe transmission and reflection calculated from the outgoing fields as a function of pump-probe time delay are shown in Fig~\ref{fig:pp1}.
In each case, we adjusted $\eta$ to match the cross polarized data and used this value for the copolarized data.
We found $\tau = 82$ fs best fits the cross polarized transmission data and use this value of $\tau$ for all calculations.
The best fit to the model corresponds to a maximum temperature rise $\Delta T_e \approx 300$ K.
Considering the simplicity of the model the agreement with experiment is very good, suggesting that the enhancement in nonlinear interaction for co-polarized, degenerate pulses is entirely due to TBC.

At $0^\circ$ incident angle, the one-beam effective nonlinear response from electron heating is completely isotropic, since the sample is an amorphous film on a glass substrate.
TBC for an effective isotropic nonlinearity depends on polarization because it works by coherent diffraction of the pump light, which is polarized, into the outgoing probe beam.
This phenomenon has been observed in other systems with effective nonlinearities produced by free electrons.
In gases, field ionization leads to an intensity dependent free carrier density, which is also isotropic, yet this leads to a transient birefringence in degenerate pump-probe experiments \cite{wahlstrand_effect_2011}.
In plasmas, polarization-dependent TBC has been used to create high intensity polarization modulators \cite{michel_dynamic_2014}.

\begin{figure}
   \centering
   \includegraphics[width=7.5cm]{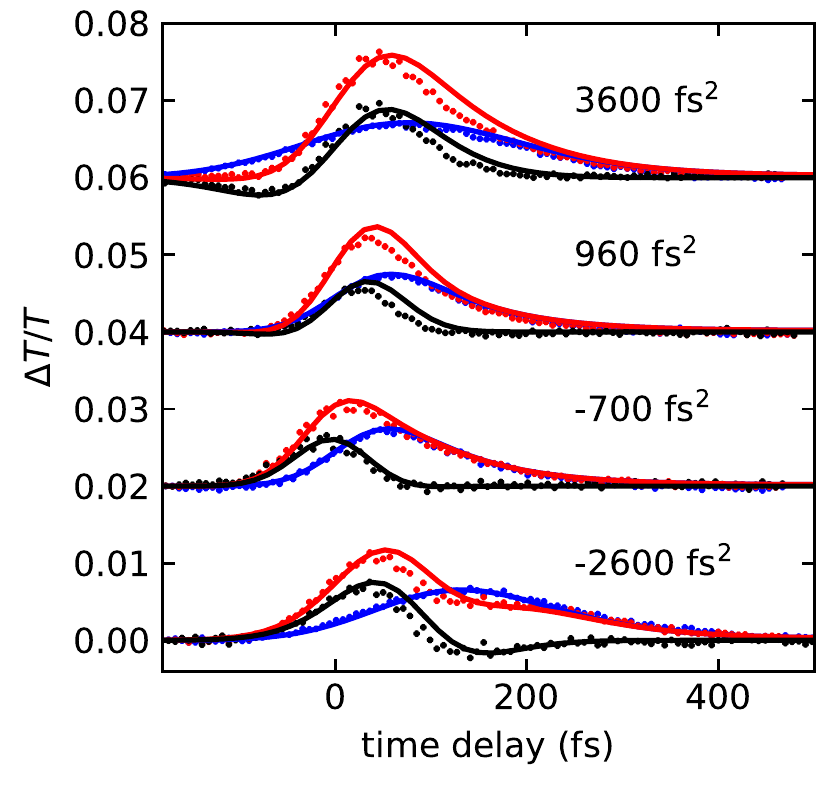}
   \caption{Experimental pump-probe traces and theory for chirped pulses at $0^\circ$ incidence. Pump probe data (dots) and theoretical curves (solid lines) for four values of the group delay dispersion. Experimental data and theory for parallel polarization are shown in red, for perpendicular in blue, and the polarization anisotropy is shown in black.
   }
   \label{chirp_fig}
\end{figure}

In other media with a delayed effective nonlinearity, TBC is sensitive to pulse chirp \cite{smolorz_femtosecond_2000,wahlstrand_effect_2013,dogariu_purely_1997}, so we next explored this in ITO.
The experiment was performed on the same ITO sample at 0$^\circ$ incidence angle.
We modified the chirp of both pump and probe pulses by inserting dispersive material (ZnSe) at the output of the OPA.
Results are shown in Fig.~\ref{chirp_fig} at four values of group delay dispersion: $-2600$ fs$^2$, $-700$ fs$^2$, 960 fs$^2$, and 3600 fs$^2$.
The curves have the same color scheme as Fig.~\ref{fig:pp1}: blue for perpendicular polarization, red for parallel polarization, and black for the polarization anisotropy.
Calculations, shown as solid lines in Fig.~\ref{chirp_fig}, reproduce the shape of the transient well if we reduce the model-predicted value of the $n_2$ coefficient by a factor of $\sim 2$.
Using the sample thickness $316$ nm and fitting the model to the experimental results in Figs.~\ref{fig:pp1} and \ref{chirp_fig}, we find $\beta = -350$ cm/GW and $n_2 = 0.013$ cm$^2$/GW.

\begin{figure}[t]
    \centering
    \includegraphics[width=7.5cm]{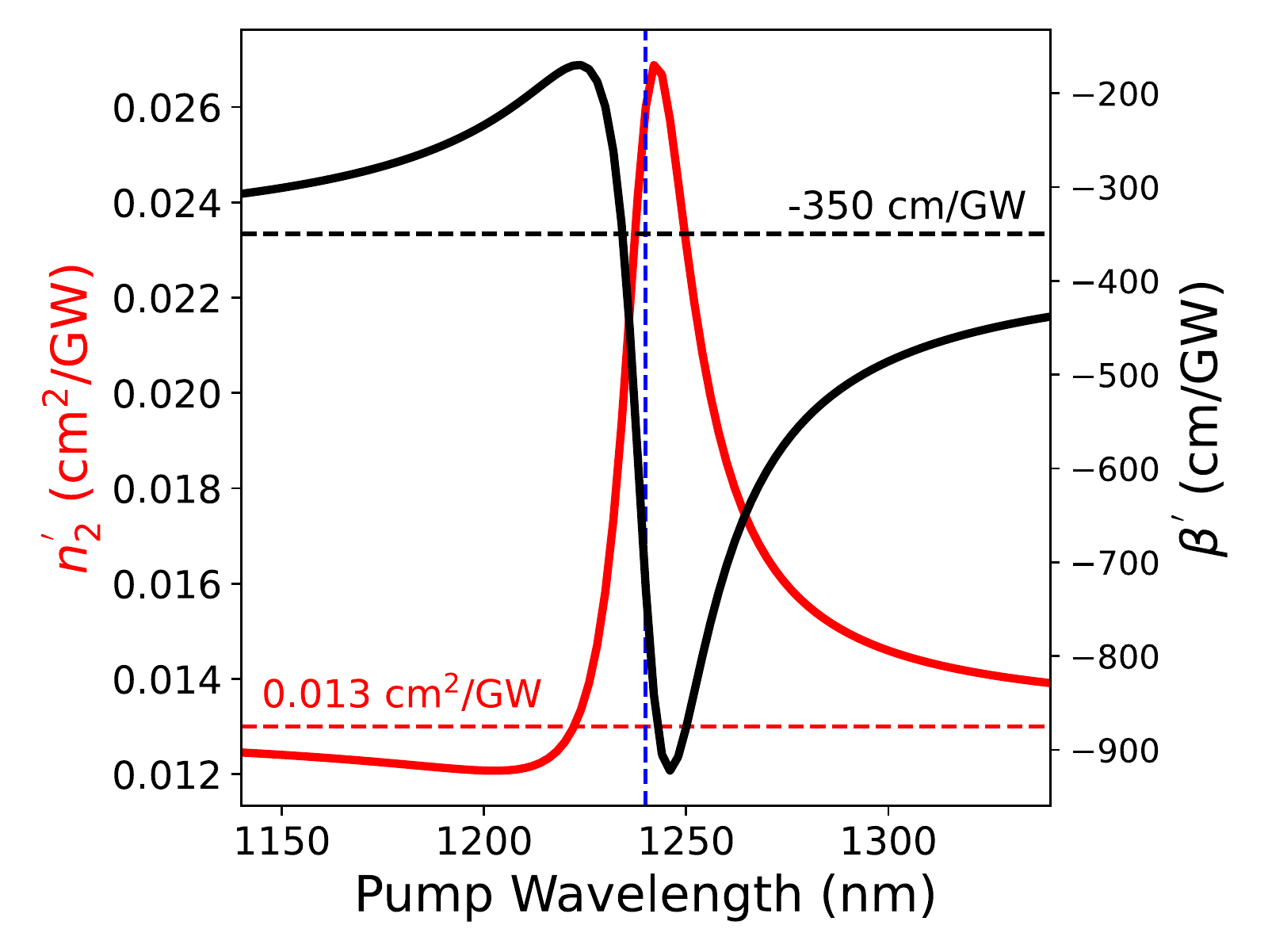}
    \caption{Nonlinear coefficients $\beta^\prime$ and $n_2^\prime$ calculated from Eqs.~(\ref{nondegen1},\ref{nondegen2}) for nondegenerate quasi-cw beams with probe wavelength 1240 nm, assuming single beam nonlinear coefficients $\beta = -350$ cm/GW and $n_2 = 0.013$ cm$^2$/GW.}
    \label{nondeg_fig}
\end{figure}

To gain more insight into the chirp dependence, it is useful to consider the limit of continuous wave beams, where analytical expressions can be derived.
The case of continuous wave beams and a purely refractive nonlinearity is treated in \cite{boyd_nonlinear_2008}.
The time dependence of the hot electron nonlinearity in ITO and similar materials is nearly identical to the Debye relaxation nonlinearity considered here, which was initially developed for photorefractive media \cite{silberberg_instabilities_1982}.
However, the relaxation time $\tau$ is many orders of magnitude faster, and both $n_2$ and $\beta$ are non-zero.
For two continuous wave beams with carrier frequencies near $\omega$, separated by detuning $\delta = \omega_{\mathrm{pump}}-\omega_{\mathrm{probe}}$, we can, using the same approach as in \cite{boyd_nonlinear_2008}, calculate modified coefficients $n_2^\prime$ and $\beta^\prime$ for the effect of one beam on the other,
\begin{eqnarray}
n_2^\prime &=& n_2 \left(1+\frac{1}{1+\delta^2 \tau^2} \right)- \beta \frac{c}{\omega} \frac{ \delta\tau}{1+\delta^2\tau^2} \label{nondegen1} \\
\beta^\prime &=&  \beta \left(1+\frac{1}{1+\delta^2 \tau^2} \right)+n_2 \frac{\omega}{c} \frac{\delta \tau}{1+\delta^2\tau^2}
\label{nondegen2}
\end{eqnarray}
See the supplemental document for more discussion.
Note that we have assumed that the coefficients do not change with $\delta$, which is possibly violated for wavelengths near the ENZ point and p polarized light at high incident angle, where $\beta$ is resonantly enhanced and $n_2$ can change sign \cite{caspani_enhanced_2016}.
Depending on $\delta$, $n_2'$ or $\beta'$ can be reduced or enhanced with respect to $n_2$ or $\beta$.
Equations (\ref{nondegen1},\ref{nondegen2}) are plotted in Fig.~\ref{nondeg_fig} for probe wavelength 1240 nm and pump wavelength varied from 1140-1340 nm, using $\tau = 82$ fs, and the values of $n_2$ and $\beta$ used to fit the data in Fig.~3.

The theoretical expression for $\beta'$ is consistent with our experimental results. Time delayed, quadratically chirped pulses have a frequency difference $\delta$ that varies linearly with time, so a change in the nonlinear response with $\delta$ shows up as a change in the temporal shape of the TBC transient.
The polarization anisotropy, plotted in black in Fig.~\ref{chirp_fig}, isolates the grating contribution $\beta'-\beta$, so a negative value corresponds to a situation where the total absorptive nonlinearity $\beta'$ is smaller than the smooth (one beam) absorptive nonlinearity $\beta$.
As can be seen in Fig. \ref{nondeg_fig}, this condition occurs when the probe wavelength is shorter than the pump wavelength.

At a pump wavelength of $1240$ nm, where the pump is degenerate with the probe ($\delta = 0$), each nonlinear coefficient is enhanced by a factor of 2.
On either side of this degeneracy, the absorptive and refractive components of the nonlinearity are mixed, resulting in an increased nonlinearity at a slightly longer wavelength than 1240 nm.
The size of the modification of the nonlinearity for nondegenerate pulses depends on the ratio of the underlying nonlinearities $n_2\omega/c\beta$.
For $|n_2\omega/(c\beta)| > 2^{3/2}$, there exists a value of $\delta$ where the modified two-photon absorption coefficient $\beta^\prime = 0$.
For $|n_2\omega/(c\beta)| < 2^{-3/2}$, there exists a value of $\delta$ where the modified Kerr coefficient $n_2^\prime = 0$.
In potential applications of the enhanced nonlinearity in materials like ITO, such as all-optical modulation \cite{yu_all-optical_2016,wang_monolithic_2019} and photonic neural networks \cite{miscuglio_all-optical_2018,zuo_all-optical_2019}, TBC could be used to tailor the interaction.

We have shown that two-beam coupling enables a new degree of tailoring the all-optical nonlinearity in hot-electron materials.
Indeed, we find that the ENZ-enhanced nonlinearity of ITO can be significantly altered both temporally and spatially by nonlinear coupling with another beam.
At degeneracy, two-beam coupling results in an enhancement of a factor of 2, while just just off the degeneracy point, greater enhancement is possible.
More importantly, we find that the absorptive and refractive nonlinearities are mixed, which enables sophisticated tailoring of the nonlinearity.
The enhancement and/or modification of the nonlinearity from all-optical beam interaction can be a powerful tool for ultrafast optics applications, enabling capabilities for applications in all-optical signal processing such as neural networks \cite{miscuglio_all-optical_2018}, active nanophotonic and plasmonic devices, optical limiters, and advanced spectroscopy.

\begin{acknowledgements}V.S. acknowledges support by the Multidisciplinary University Research Initiative (MURI) program (FA 9550-17-1-0071) through the Air Force Office of Scientific Research (AFOSR).
\end{acknowledgements}

\appendix

\section{Experiment Details}
The optical parametric amplifier was pumped by a Yb-doped solid state laser (Light Conversion Carbide 40 and Orpheus \cite{blurb})
For the experiments, the probe is nearly normally incident on the sample, and the angle between the pump and probe beams is 10$^\circ$.
The differential transmission and reflection were measured using a mechanical chopper and lock-in amplifier.
The pulse duration, measured by cross correlation of the pump and probe pulses in a nonlinear crystal, is 67 fs full width at half maximum, assuming Gaussian pulses.

\begin{figure}
   \centering
   \includegraphics[width=9cm]{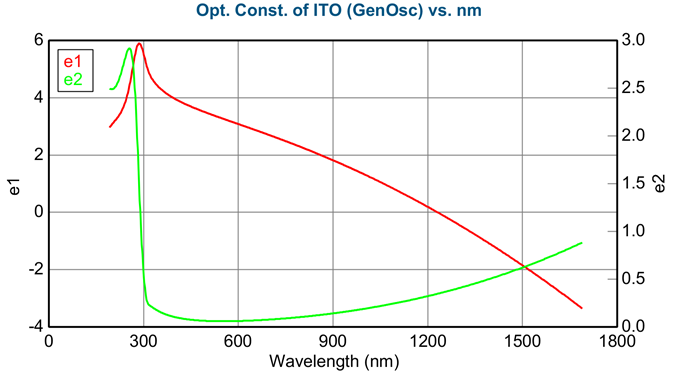}
   \caption{ITO sample dispersion measured using spectroscopic ellipsometry.
   }
   \label{sample_fig}
\end{figure}

The sample is a commercial indium tin oxide (ITO) film.
The substrate material used is fused silica, which has a relatively small, purely refractive Kerr nonlinearity at near infrared wavelengths.
The optical properties of the sample, measured using spectroscopic ellipsometry, are shown in Fig.~\ref{sample_fig}.
The sample thickness of 316 nm and epsilon near zero (ENZ) wavelength of 1240 nm were determined from the ellipsometry results.

\section{Theory for Pulses}

As described in the main text, the equations for temperature are $C_e dT_e/dt =-g_{el} (T_e - T_l) + S(t)$, and $C_l dT_l/dt = g_{el} (T_e - T_l)$.
For $C_e \ll C_l$, the time dependence of the response is not sensitive to the precise values of $C_e$ and $C_l$, so we use $C_l/C_e = 100$.
We also assume the heating term is linear in the intensity, which is a good approximation at the relatively low intensities used here, but would fail at higher intensity, where the heating saturates \cite{reshef_beyond_2017,wang_extended_2019}.
To calculate two-beam coupling, we calculate the time-dependent electron temperature distribution inside the pump spot, including probe interference.
We find $T^s_e(t)$ by solving the equations for electron and lattice temperature for $S(t) = \eta I_e(t)$.
To calculate the interference term $T^g_e(t)$ we solve the equations for temperature for $S(t) = \eta I_e(t) + h n_0 c/(8\pi) A_e^*(t) A_x(t-t_d) e^{i\delta t}$ and then subtract $T^g_e(t)$ to isolate the grating term.
Note that $T^g_e(t)$ is a complex quantity because it carries the relative phase of the pump and probe envelopes.
This phase drops out of the final calculation.

To calculate the complex dielectric function from $T_e$, we use the modified Drude model developed by Wang \emph{et al.} for ITO \cite{wang_extended_2019},
\begin{equation}
\varepsilon(\omega,T_e) = \varepsilon_\infty - \frac{Ne^2}{\varepsilon_0[m^*(T_e)\omega^2 +ie\omega/\mu(T_e)]},
\label{modDrude}
\end{equation}
where $\varepsilon_\infty$ is the high frequency permittivity, $N$ is the carrier density, $e$ the electron charge, $\varepsilon_0$ the vacuum permittivity, and $m^*(T_e)$ and $\mu(T_e)$ the electron effective mass and mobility, which vary with electron temperature.
In this model the plasma frequency $\omega_p = Ne^2/[\varepsilon_0 m^*(T_e)]$ decreases with $T_e$ and the scattering rate $\gamma = e/[m^*(T_e) \mu(T_e)]$ increases.
For our calculations, we use the parameters reported in \cite{wang_extended_2019} with one important modification:
at the relatively low $T_e$ relevant to our experiment, the model predicts a significant effective higher order response that is positive, whereas at higher $T_e$ the response saturates (i.e. higher order terms are negative) \cite{reshef_beyond_2017,wang_extended_2019}.
This behavior is inconsistent with the linear power dependence we observe at the low intensities accessible in our experiment.
We therefore modified the model, forcing it to be linear at low intensity with the same average slope as the full model in the range of approximate linearity at higher intensity.
For more details, see the next section.

Complex transmission and reflection coefficients $t^s$, $t^g$, $r^s$, and $r^g$ are found using a standard transfer matrix calculation.
The outgoing probe envelope in the forward direction is given by
\begin{eqnarray}
A^{\mathrm{trans}}_p(t) &=& t^s(t) A_p(t-t_d) + t^g(t) A_e (t) e^{i\delta t}, \\
A^{\mathrm{refl}}_p(t) &=& r^s(t) A_p(t-t_d) + r^g(t) A_e (t) e^{i\delta t}.
\end{eqnarray}
The first term in each equation above is the time-dependent dielectric constant due solely to the pump-induced change in optical properties.
The second term, which is only non-zero for parallel polarization, is caused by pump light diffracted into the probe direction.
Including a small smooth term proportional to the lattice temperature improved the fits at large time delays.
The value of $\tau$ that best fits the data varies between 82 fs and 130 fs, depending on whether the geometry was transmission or reflection and the incident angle.
We speculate that this may be caused by a depth dependent temperature not included in our modeling.

\section{Calculating the dependence of optical properties on electron temperature}

\begin{figure}[t]
   \centering
   \includegraphics[width=9cm]{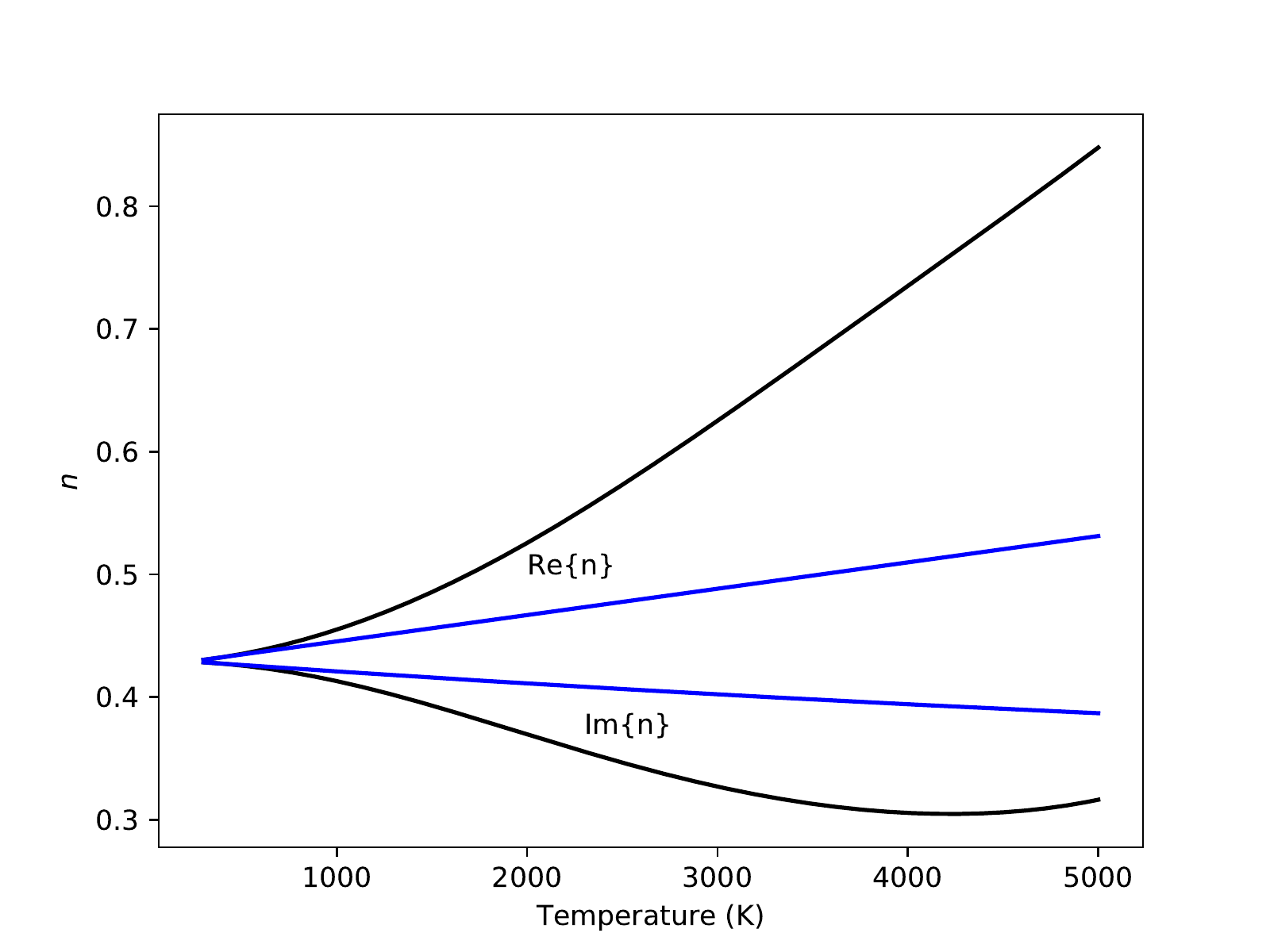}
   \caption{Calculated real and imaginary parts of the refractive index as a function of electron temperature. Black: Wang \emph{et al.} parameters \cite{wang_extended_2019}. Blue: Linearized model, which matches the slope at lowest intensity.
   }
   \label{linearization_fig}
\end{figure}

We base our calculation of the optical properties of ITO as a function of electron temperature on the model developed by Wang \emph{et al.} \cite{wang_extended_2019}, as described in the main text.
They reported parameters for the commercial ITO film studied: $N=1.5\times 10^{21}$ cm$^{-3}$, $m^* = 0.3964 m_e$, $\varepsilon_\infty = 3.404$, $E_F = 0.8793$ eV (Fermi level at zero temperature), and $C=0.4191$ eV$^{-1}$ (nonparabolicity parameter).
The mobility versus temperature is $\mu(T) = 18.3 + 2.13 \times 10^{-5} T^{1.53}$ in cm$^2$/(V $\cdot$ s).
The real and imaginary components of the refractive index calculated from the model are plotted in Fig.~\ref{linearization_fig} in black.

At the relatively low electron temperatures ($<700$ K) accessible in our experiment, the model is nonlinear, with a positive higher order component.
The electron temperature as a function of laser intensity is nonlinear because the heat capacity of the electron gas increases with increasing temperature, according to \cite{guo_ultrafast_2016}.
This counteracts the nonlinearity in the optical properties, resulting in an approximately linear dependence of the nonlinear refractive index with pulse intensity (see Fig.~4a in \cite{wang_extended_2019}).
Our simple model of heating assumes a linear dependence of electron temperature on pulse intensity, and the nonlinear dependence of the nonlinear refractive index from the full model resulted in a prediction that the change in transmission due to the pump pulse is nonlinear in pump intensity.
This is inconsistent with the linear dependence of the transmission and reflection change on pump intensity that we observed in the experiment.
It also produced a drastic overestimate of the two-beam coupling transient for co-polarized beams (the ``grating'' signal in Fig. 2c), because this signal is proportional to $d\alpha/dI$ \cite{wahlstrand_effect_2011}.
As a result, we adjusted the model to have a linear dependence on temperature matching the low intensity slope.
This linearized version of the model is shown in Fig.~\ref{linearization_fig} in blue.
We emphasize that changing the actual slopes of the curves would not change the conclusion of the paper.
We adjusted the phenomenological heating parameter $\eta$ to fit the cross polarized data.
The fits to the experimental data are good as long as the change in optical properties is linear in the pump fluence.

\section{Two-beam coupling in the continuous wave limit}
A derivation of Eqs.~(2,3) in the main text is provided here.
We closely follow the approach in \cite{boyd_nonlinear_2008}, section 7.4, which was in turn based on \cite{silberberg_instabilities_1982,silberberg_optical_1984}.
The exponentially decaying nonlinearity used is referred to as "Debye relaxation" and was developed to model the nonlinearity in a  photorefractive material.
It happens to be suitable to model the nonlinearity in ITO and related materials, except that we must include an absorptive nonlinearity in addition to the Kerr coefficient $n_2$.
For this we define a complex nonlinear coefficient $\nu = n_2 + i\beta/k$, where $\beta$ is the two-photon absorption coefficient and $k=\omega/c$.
The nonlinear refractive index change versus time obeys (see Eq.~(7.4.7) in \cite{boyd_nonlinear_2008})
\begin{equation}
\tau \frac{d \Delta n}{dt} + \Delta n = \nu I,
\end{equation}
and it can be easily seen that this is approximately true for the two-temperature model, as long as we neglect the lattice temperature's effect on the optical properties.
We use the observed cooling time $\tau \approx 85$ fs.

The assumptions behind the equations above also break down in the saturation regime \cite{reshef_beyond_2017}.

We proceed exactly as described in \cite{boyd_nonlinear_2008}.
We assume two beams with central frequencies $\omega_1$ and $\omega_2$, which are sufficiently closely spaced that we can use $\omega = (\omega_1 + \omega_2)/2$ wherever the laser frequency appears.
As mentioned in the main text, we are assuming that $n_2$ and $\beta$ are the same at $\omega_1$ and $\omega_2$.
This holds as long as $\omega_1-\omega_2$ is relatively small.
Defining $\delta = \omega_1-\omega_2$, the equation for the propagation of beam 2 is (Eq.~(7.4.15))
\begin{equation}
\frac{dA_2}{dz} = 2in_0 \nu \frac{\omega}{c} \left[ \left(|A_1|^2+|A_2|^2 \right) A_2 + \frac{|A_1|^2 A_2}{1+i\delta \tau} \right],
\end{equation}
This complex coefficient includes both nonlinear refraction and absorption.
To derive the gain or loss, introduce intensities
\begin{equation}
I_1 = 2n_0 \epsilon_0 c A_1 A_1^* \mathrm{~and~} I_2 = 2n_0 \epsilon_0 c A_2 A_2^*,
\end{equation}
and we see that
\begin{equation}
\frac{dI_2}{dz} = 2n_0 \epsilon_0 c \left( A_2^* \frac{dA_2}{dz} + A_2 \frac{A_2^*}{dz} \right).
\end{equation}
This leads to
\begin{equation}
\frac{dI_2}{dz} = \frac{2n_2\omega}{c} \frac{\delta \tau}{1+\delta^2 \tau^2} I_1 I_2 - \beta I_2^2 - \beta I_1 I_2 - \beta \frac{I_1 I_2}{1+\delta^2\tau^2}.
\end{equation}
The first term above is the conversion of the refractive nonlinearity to absorption through two-beam coupling.
The second term corresponds to two-photon absorption of the probe beam alone.
The third term corresponds to two-photon absorption where one photon comes from the probe and the other comes from the pump.
The fourth term is a reduction in the probe absorption caused by two-beam coupling.

Now consider the phase shift of one beam. Assuming $A_2 = \bar{A}_2 e^{i\phi}$, we find
\begin{equation}
\frac{d\phi_2}{dz} = -2\beta \frac{\delta \tau}{1+\delta^2 \tau^2} I_1 + \frac{\omega}{c} \left( n_2 I_2 + n_2 I_1 + n_2 \frac{I_1}{1+\delta^2\tau^2} \right).
\end{equation}
The first term above is the conversion of the absorptive nonlinearity to phase modulation through two-beam coupling.
The second term (first in parentheses) corresponds to self phase modulation of the probe beam.
The third term corresponds to cross phase modulation of the probe beam by the pump beam.
The fourth term is a reduction in the self phase modulation caused by two-beam coupling.

By inspection, we see that we can define modified coefficients given by Eqs.~(2,3).


\begin{thebibliography}{10}
\bibitem{blurb}
The identification of commercial products in this paper is for information only and does not imply recommendation or endorsement by the National Institute of Standards and Technology.
\end{thebibliography}
\end{document}